\documentclass[preprintnumbers,amsmath,amssymb,twocolumn]{revtex4}
\usepackage[dvips,colorlinks=true,citecolor=red,filecolor=green,linkcolor=blue,pdfnewwindow=true]{hyperref}
\usepackage{graphicx}
\usepackage{makeidx}
\usepackage{bm}
\usepackage[title]{appendix}

\begin{document}
\title{Bifurcation in cell cycle dynamics regulated by $p53$}
\author{Md. Jahoor Alam$^1$}
\author{Sanjay Kumar$^2$}
\author{Vikram Singh$^3$}
\author{R.K. Brojen Singh$^{1}$}
\email{brojen@mail.jnu.ac.in}
\affiliation{$^1$School of Computational and Integrative Sciences, Jawaharlal Nehru University, New Delhi-110067, India. \\
$^2$ Department of Computer Science, Jamia Millia Islamia,New Delhi 110025, India.\\ $^3$School of Life Sciences, Central University of Himachal Pradesh, Dharamshala-176215, India.}

\begin{abstract}
We study the regulating mechanism of $p53$ on the properties of cell cycle dynamics in the light of the proposed model of interacting $p53$ and cell cycle networks via $p53$. Irradiation ($IR$) introduce to $p53$ compel $p53$ dynamics to suffer different phases, namely oscillating and oscillation death (stabilized) phases. The $IR$ induced $p53$ dynamics undergo collapse of oscillation with collapse time $\Delta t$ which depends on $IR$ strength. The stress $p53$ via $IR$ drive cell cycle molecular species $MPF$ and cyclin dynamics to different states, manely, oscillation death, oscillations of periods, chaotic and sustain oscillation in their bifurcation diagram. We predict that there could be a critical $\Delta t_c$ induced by $p53$ via $IR_c$, where, if $\Delta t\langle\Delta t_c$ the cell cycle may come back to normal state, otherwise it will go to cell cycle arrest (apoptosis).
\end{abstract}

\maketitle

\section{Introduction}

$p53$ is well known for its abnormally long stability in response to the stress available against genomic integrity \cite{lane}. It conglomerated with its negative inhibitor $MDM2$ in the nucleus due to their strong interaction \cite{men}. When the cell is in stress condition (due to irradiation, stress inducer molecule etc), $p53$ concentration level rises which leads to cell cycle arrest until repair or doctoring takes place of the impaired DNA. If the repair is not successful the system move towards the apoptosis \cite{mic,bar,vog,vou}. The transcriptional ability of the $p53$ is kept under controlled level at normal state due to its negative feedback interaction with $MDM2$ \cite{mom}. The hyperbolized concentration of $MMD2$ helps in degradation of the $p53$ protein because of its E3-ligase activity, causing adherence of ubiquitin to the lysine rich C-terminal of the $p53$ molecule \cite{mak,hon,sch}. Introduction of stress in the system is sensed by the activation of $ARF$ protein, initially situated in nucleolar region in the form of nucleophosmin shifts to the nucleoplasm in its independent and active cast, to mark $MDM2$ for its degradation, thus assisting the $p53$ stability \cite{pom,hon1,mid}. Triggering of $p53$ in response to stress leads to the expression of several downstream genes apart from the $MDM2$. 

$p21$ protein is one of the most important proteins which is found to be expressed due to $p53$ accumulation in the cell \cite{dei}. $p53$ acts as a transcription factor for $p21$. It is also reported that $p21$ expression is directly proportional to the level of $p53$ in the system \cite{gal}. The role of $p21$ in controlling G1 phase checkpoint has been widely studied but its role in controlling G2 phase checkpoint is comapratively less studied \cite{bun,bat,agu}. The G2 phase checkpoint interuption leads to the disruption of cell cycle that leads to halt mitosis \cite{dei}. The cyclin-cdk interaction leads to the formation of $MPF$ (Maturation Promoting Factor) \cite{har}. The formation of $MPF$ is very important for transition of G2 phase to mitosis phase \cite{apre}. The $p21$ protein is reported as antagonist for the formation of $MPF$. Several experimental results suggests that $p21$ directly interacts with $cdk$ and also with cyclin leading to the inhibition of both $cdk$ as well as cyclin \cite{fun}. It is also reported that the interaction of $cdk$ and $p21$ causes to halt in DNA replication \cite{sor,apre}.
\begin{figure*}
\label{fig1}
\begin{center}
\includegraphics[height=270 pt, width=450 pt]{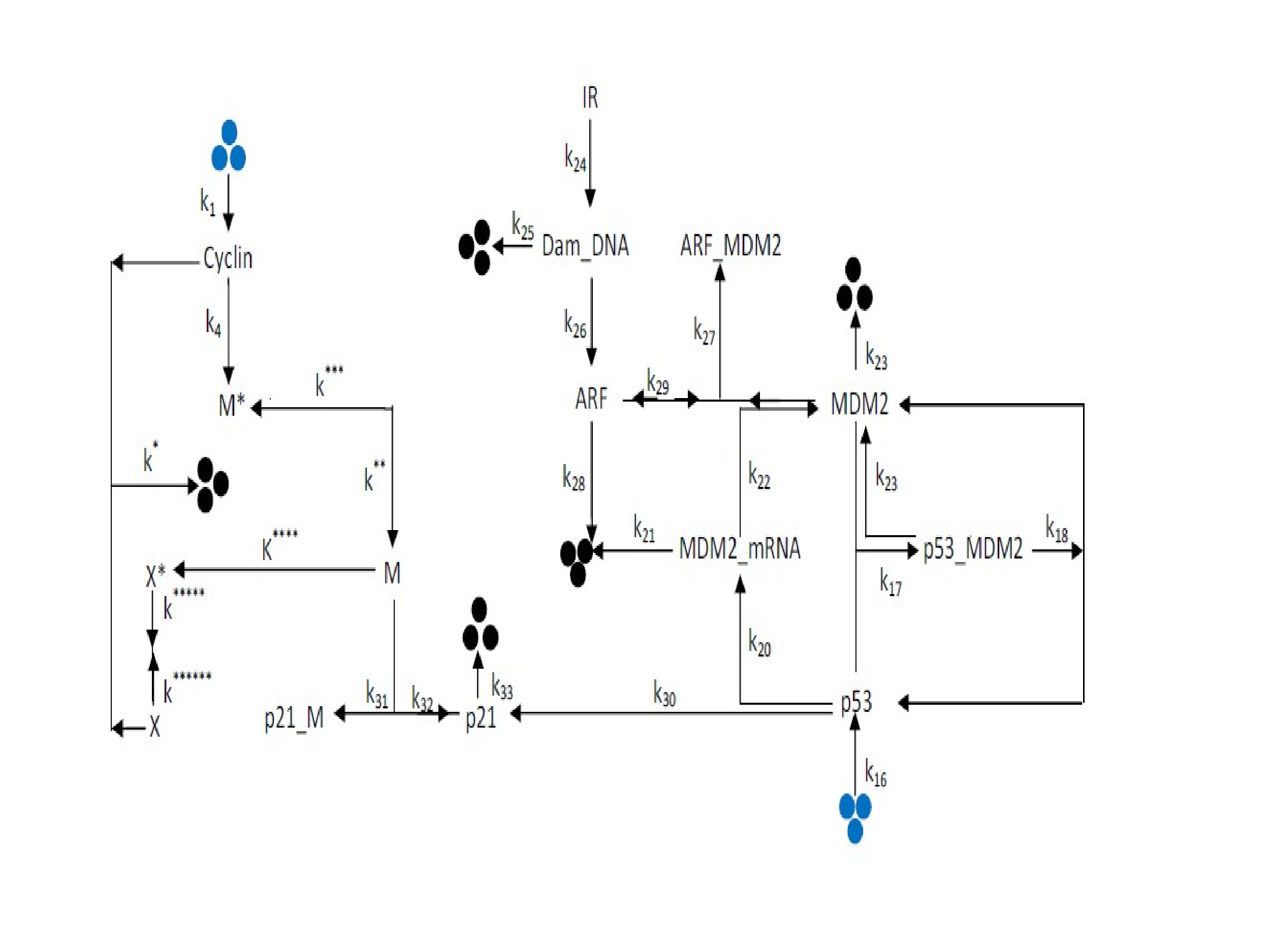}
\caption{The schematic diagram of interaction of p53-Mdm2 reaction network cell cycle oscillator.}
\end{center}
\end{figure*}

Cyclin, in cell cycle process, is an important protein which interacts with cyclin dependent kinases and forms $MPF$. The $MPF$ is responsible for the activation of pRb (Retinoblastoma protein), and helps the liberation of transcription factor $E2F$ from its inhibitory. This $E2F$ maintains the expression profile of genes required to ingress the S-phase of the cell division cycle \cite{res,hin,lam}. Further, it is reported by several experimental results that $p21$ can directly interact with $MPF$ and forms complex and then dissociate \cite{bun,agu}. Hence, $p53$ can able to cross talk with $MPF$ and cyclin through $p21$.

We focus in this work to study and find out the behaviour of different molecular species which are actively involved in the checking of cell cycle at G2 phase regulated by $p53$. We proposed an integrated model of $p53$ and cell cycle network to find out the impact of $p53$ regulator on cell cycle via $p21$ protein. We organized our work as follows. We explained our proposed model in section II. The results of the large scale simulation of the model is given in section III with discussion. The conclusion based on our results is provided in section IV.

\section{Model of cell cycle regulated by $p53$}

We present a model which brings together $p53-MDM2$ regulatory network \cite{pro} and cell cyle \cite{gold} via $p21$ protein (Fig. 1) in the light of various theoretical and experimental reports. The model is described briefly as follows. The main component of $p53-MDM2$ regulatory network is the feedback loop between $p53$ and $MDM2$ \cite{pro}. $p53$ and $MDM2$ interact to form $p53-MDM2$ complex with a rate constant $k_{17}$ \cite{pro}, followed by dissociation of the complex to its respective components with a rate costant $k_{18}$ \cite{moll,jah}. The transcription rate of $MDM2$ gene to its $mRNA$ ($MDM2-mRNA$) is takes place with rate constant $k_{20}$, followed by translation of $MDM2-mRNA$ to $MDM2$ with a rate constant $k_{22}$ \cite{pro,fin} and its ($MDM2-mRNA$) self-degradation with a rate constant $k_{21}$ \cite{lah}. The ubiqiutination of $MDM2$ protein occurs with rate constant $k_{23}$. The $p53$ synthesis is taken placed with a rate constant $k_{16}$, and gets ubiquitinized at the rate constant $k_{19}$ \cite{fin}. The DNA damage in system is introduced via irradiation with an estimated rate constant of $k_{24}$ \cite{pro}. The repair of the DNA damage is then occured at a rate constant $k_{25}$ \cite{vil,schu}. The activation of $ARF$ due to DNA damage takes place at a rate constant $k_{26}$\cite{pro}. Further, this activated form of $ARF$ interact with $MDM2$ protein and forms $ARF-MDM2$ complex with a rate constant $k_{27}$ \cite{khan}. The degradation of $ARF$ protein is reported to occur at a rate constant $k_{28}$ \cite{kuo}. $ARF$ based degradation of the $MDM2$ takes place by getting targeted to the complex via proteosome recognition with a much faster rate constant $k_{29}$ than individual degradation rates \cite{zhan}. The $p53$, being a transcription promoting factor for many of the proteins, also transcribes the gene responsible for the manufacture of $p21$ protein with a rate constant $k_{30}$ as presumed by the approximations made to attain the appropriate oscillations and arrests \cite{dei,xio,deir}. 

The $p21$ protein is capable of making complex with the cell division promoting factor $MPF$ with a rate constant $k_{31}$ \cite{bun,agu} with respect to the amount of concerned molecules present in the system \cite{har}. Then the inhibition of $MPF$, or more appropriately G2 associated $Cyclin-Cdk$ complex, by $p21$ is approximated with a rate constant $k_{32}$ \cite{xio,wag}. $p21$ then gets degraded by the virtue of its half life in the system with a rate constant $k_{33}$ \cite{agu,res}. The Cyclin is assumed to translate at the rate constant $k_1$\cite{morl}. Further, ubiquitin dependent Cyclin degradation or protease independent degradation of the Cyclin is reported to happen at a rate constant $k^{*}$ \cite{stra}. The degradation of the Cyclin due to effect of protease activation during cyclin accumulation and interaction between inactive form of $MPF$ with Cyclin takes place with a rate constant $k_4$ \cite{gold}. Formation of activated form of $MPF$ (M) occurs due to interaction of cyclin with inactive $MPF$ ($M^{*}$) with a rate constant $k^{**}$\cite{gold,murr2,mins,rom}. Further this activated form of $MPF$ (M) converts to inactivated form ($M^{*}$)with a rate constant $k^{***}$ \cite{murr2,rom}. The activated form of $MPF$(M) interact with inactive protease($X^{*}$) to generate activated form of protease (X) with a rate constant $k^{****}$ \cite{gold,mins1,bue}. The generation of activated form of cyclin protease ($X$) occurs due to interaction of cyclin protease with inactive $X^{*}$ with a rate $k^{*****}$\cite{gold,murr2}. The activated form of protease ($X$) can convert into inactive form ($X^{*}$) with a rate constant $k^{******}$\cite{gold,rom}. The lists of molecular species and biochemical reaction channels involved in this proposed model are listed in Table 1 and Table 2 respectively. 

\begin{figure}
\label{fig2}
\begin{center}
\includegraphics[height=250 pt]{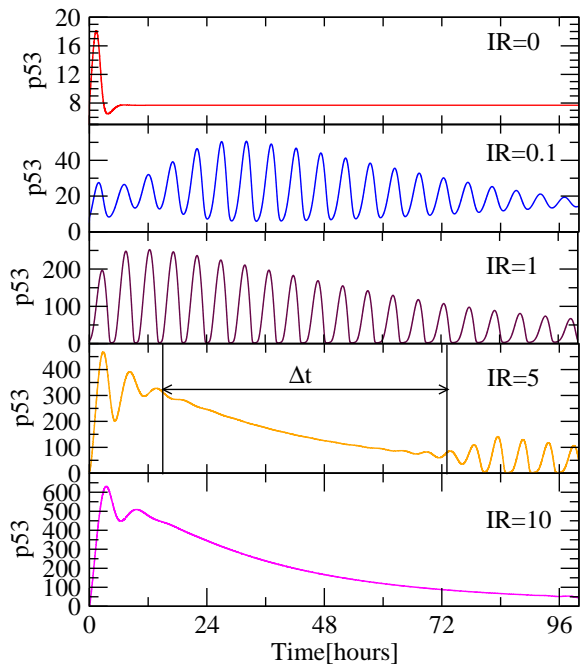}
\caption{Plot shows the variation in the concentration and oscillatory pattern of p53 protein due to the effect of various exposure of IR i.e (0,0.1,1,5,10).}
\end{center}
\end{figure}

The biochemical reaction network shown in Fig. 1 are represented by the twenty five reaction channels listed in Table 2, which are participated by thirteen molecular species (Table 1) defined by a vector at any instant of time $t$, ${\bf x}(t)=\{x_1(t),x_2(t),\dots,x_N(t)\}^T$, where, T is the transpose of the vector and $N=13$. The variables are the concentrations of the molecular species. The time evolution of these variables can be translated from the twenty five reaction channels into the following set of nonlinear ordinary differential equations (ODE) based on Mass action law of chemical kinetics,
\begin{eqnarray*}
\label{ma}
\frac{dx_1}{dt}&=&k_1-\frac{k_2x_1x_3}{k_3+x_1}-k_4x_1\\
\frac{dx_2}{dt}&=&\frac{k_5(1-x_2)}{x_6+(1-x_2)}-\frac{k_7x_2}{k_8x_2}-k_{31}x_{12}x_2\\
\frac{dx_3}{dt}&=&\frac{k_9(1-x_3)}{k_{10}+(1-x_3)}-\frac{k_{11}x_3}{k_{12}+x_3}\\
\frac{dx_4}{dt}&=&k_{16}+k_{18}x_6-k_{17}x_4x_5\\
\frac{dx_5}{dt}&=&k_{22}x_7+k_{19}x_6+k_{18}x_6-k_{23}x_5-k_{17}x_5x_4\\
&&-k_{27}x_5x_8\\
\frac{dx_6}{dt}&=&k_{17}x_4x_5-k_{18}x_6-k_{19}x_6\\
\frac{dx_7}{dt}&=&k_{20}x_4-k_{21}x_7\\
\frac{dx_8}{dt}&=&k_{26}x_{11}+k_{29}x_9-k_{27}x_{5}x_8-k_{28}x_8\\
\frac{dx_9}{dt}&=&k_{27}x_5x_8-k_{29}x_9\\
\frac{dx_{10}}{dt}&=&-k_{24}x_{10}\\
\frac{dx_{11}}{dt}&=&k_{24}x_{10}-k_{25}x_{11}\\
\frac{dx_{12}}{dt}&=&k_{30}x_4-k_{31}x_{2}x_{12}+k_{32}x_{13}-k_{33}x_{12}\\
\frac{dx_{13}}{dt}&=&k_{31}x_{12}x_2-k_{32}x_{13}
\end{eqnarray*}
where, the expressions for $M^*$ and $X^*$ in the Fig. 1 are given by, $M^*$=1-$x_{10}$ and $X^*$=1-$x_{11}$. The set of coupled ODEs can be solved using Runge Kutta method of standard numerical integration algorithm \cite{pre}.

\begin{figure*}
\label{fig3}
\begin{center}
\includegraphics[height=280 pt, width=350 pt]{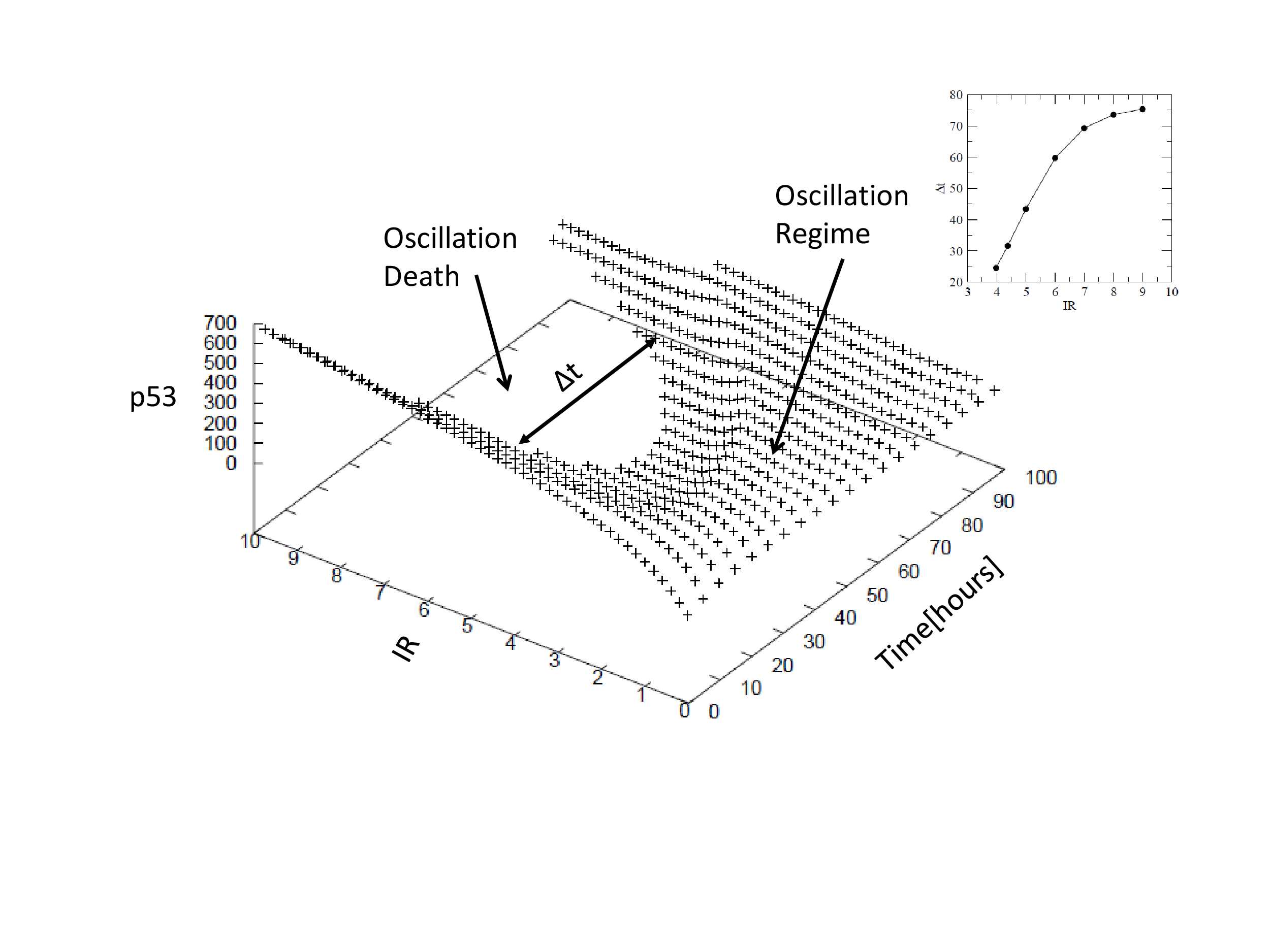}
\caption{Plot for showing the impact of IR on p53 maxima. Different p53 maxima observed at different values of IR with respect to time. The p53 maxima verses IR dose is shown at left hand side inset and also IR dose verses time is shown in right hand inset.}
\end{center}
\end{figure*}

\section{Results and discussion}

We numerically simulate the proposed model and the results demonstrate new phenomena in bifurcation diagram which may be significant to correlate with various experimental situations. The interaction of $p53$ regulatory network and cell cycle network highlights different form of signal processing between unidentical networks which could be the way of regulating one another. We study the complicated way of this interaction in order to understand some of the basic mechanisms of network interaction.

\subsection{Dynamics of $p53$ driven by irradiation}

We first present the spatio-temporal behaviour of $p53$ upon exposure of irradation in Fig. 2. The $p53$ dynamics maintains minimum concentration level at $IR=0$ (normal condition). As $IR$ dose increases $p53$ start showing damped oscillatory behaviour (Fig. 2 second and third panels) indicating stressed behaviour of $p53$. The increase in $IR$ dose induce increase in time to attain stability of $p53$ dynamics (amplitude death) indicating increase in unstability of $p53$ dynamics (Fig. 2 third panel). This could be due to the fact that the increase in $IR$ dose may cause high DNA damage leading to more stress in $p53$. 

However, if the $IR$ dose is comparatively strong ($IR=5$), the damage within the DNA is also high which may cause the collapse of the $p53$ oscillatory behaviour (Fig. 2 fourth panel) and then repaired back the DNA damage to come back to $p53$ oscillatory condition. We also found that the time of collapse ($\Delta t$) increases as $IR$ dose increases (Fig. 2 fifth panel) and it becomes difficult to repair back the DNA damage. In general $p53$ will collapse forever and will not be recovered back if $\Delta t\rightarrow\infty$ (probably case of apoptosis). However, in real situation, one probably can define a critical $\Delta t_c$ such that, if $\Delta t\langle\Delta t_c$, $p53$ could come back after DNA repair, and otherwise it will go to apoptosis. Nevertheless, it is very difficult to find out this $\Delta t_c$. 
\begin{figure*}
\label{fig4}
\begin{center}
\includegraphics[height=280 pt]{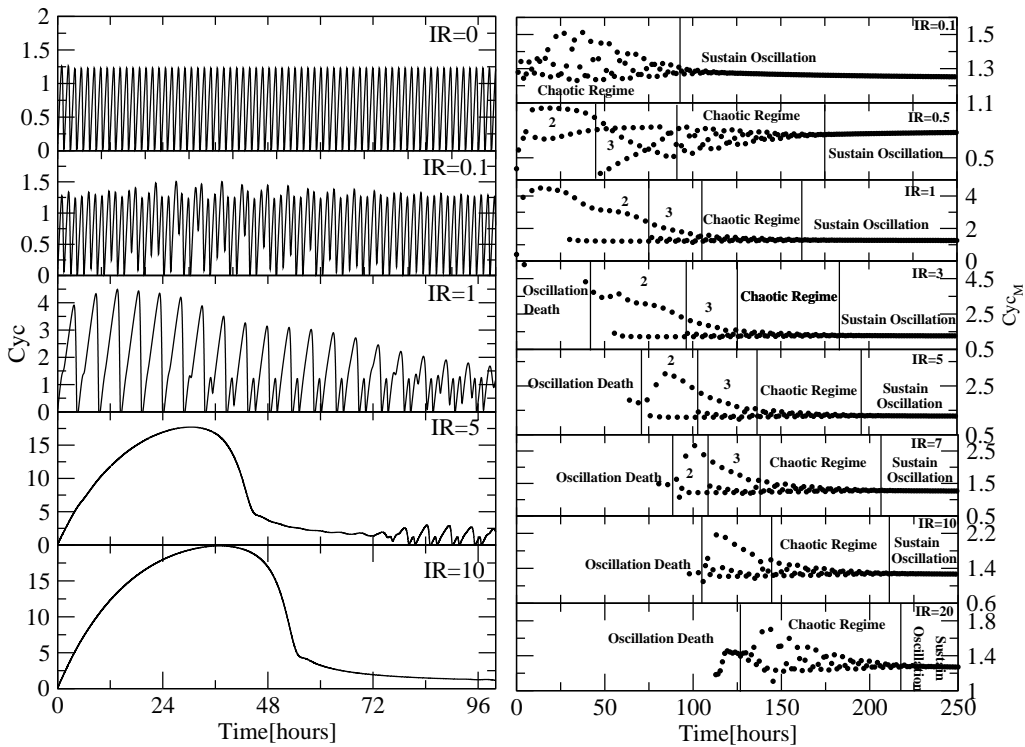}
\caption{Plot shows the variation in the oscillatory pattern of cyclin due to the effect of various exposure of IR i.e (0,0.1,1,5,10) at left side panels and their corresponding bifurcation diagram are shown at right panels.}
\end{center}
\end{figure*}

\subsection{Phase diagram of $p53$ compeled by $IR$}

We simulate the maxima of $p53$ amplitudes after removing the transients as a function of $IR$ (Fig. 3) to capture the different phases namely oscillation and oscillation death regimes. The behaviour of $\Delta t$ as a function of $IR$ follows the functional form $\Delta t=\frac{A}{B+e^{-IR}}$ with the values of $A=6778$ and $B=0.00887$ (fitting values of the function to the data) (Fig. 3 inset). The separation between two phases oscillation death and oscillating regimes are clearly visible after the $IR\sim 3.45$ and $\Delta t$ increases as $IR$ increases. 

Generally as $\Delta t\rightarrow\infty$ when $IR\rightarrow\infty$, but numerically we approximately found that after $IR=R_c\sim 11$ $\Delta t$ become $\Delta t_c\sim 79$ hours and becomes constant (Fig. 3 inset). This means that for any $\Delta t\langle\Delta t_c$, the $p53$ can able to recover back to normal stable state by repairing DNA damage, otherwise, the system can't able to come back to normal state, but will go to apoptosis.

\subsection{Bifurcation in Cyclin regulated by $p53$}

Since cell cycle and $p53$ regulatory networks are interacted through $p21$ (Fig. 1), the temporal behaviour of cyclin can be regulated by $p53$ via $IR$ and $p21$. When $IR=0$, the two networks work in normal condition, leaving $p53$ dynamics at low level (stabilized state) (Fig. 2 upper panel) and sustain oscillation in cyclin dynamics (Fig. 4 upper left panel). As $IR$ increases, $p53$ will get activated through DNA damage giving oscillatory behaviour affecting the dynamics of cyclin. When $IR=0.1$, the cyclin dynamics shows chaotic behaviour upto $t=145$ hours, and then the dynamics becomes sustain oscillation (Fig. 4 second left panel and upper right panel). The chaotic behaviour in cyclin dynamics could due to the sudden activation in $p53$ dynamics due to $IR$ irradiation.

Now as $IR$ increases ($IR=0.5$), we get various situations in the cyclin dynamics, namely, the emergence of period two (for $t\sim [10-40]$ hours), period 3 (for $t\sim [40-85]$ hours), chaotic regime (for $t\sim [85-175]$ hours) and sustain oscillation regime (for $t\rangle 175$ hours) (Fig. 4 second right upper panel). Further, as $IR$ increases the emergence of oscillation death regime started to exist in the cyclin dynamics (Fig. 4 fourth right panel onwards) and the oscillation death regime become larger. Further increase in $IR$ compels the period 2 and 3 regimes to vanish after some value of $IR$ ($IR\rangle 9$) and the chaotic regime becomes larger.

\begin{figure*}
\label{fig5}
\begin{center}
\includegraphics[height=280 pt]{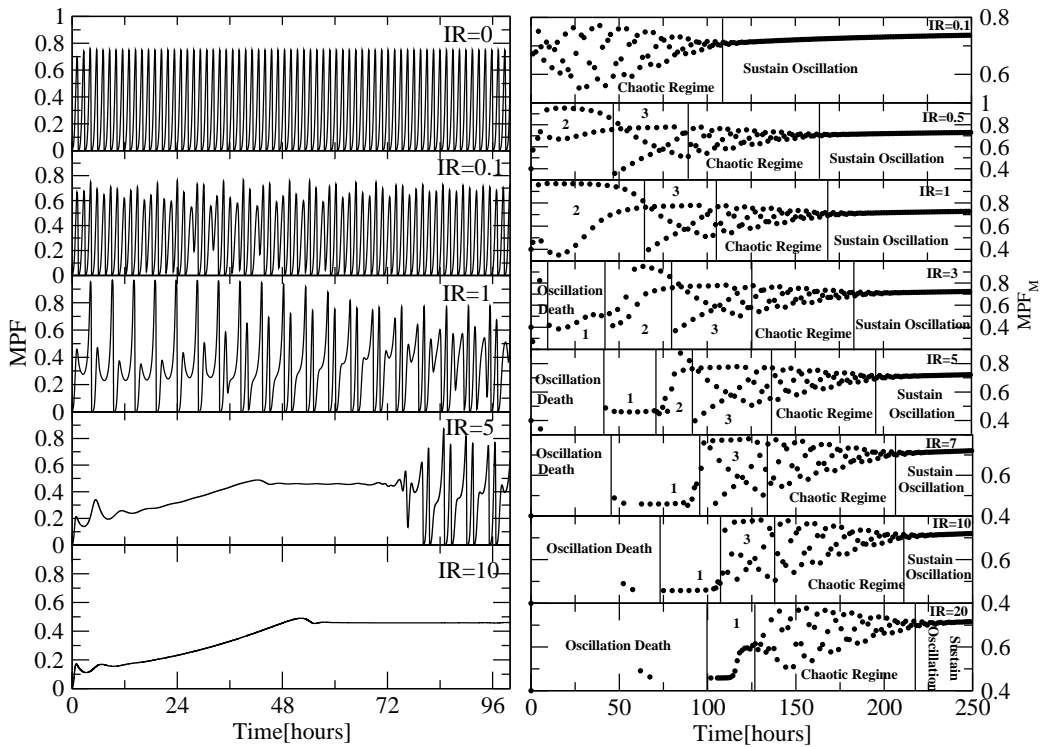}
\caption{Plot shows the variation in the oscillatory pattern of MPF(Maturation Promoting Factor) due to the effect of various exposure of IR i.e (0,0.1,1,5,10) at left side panels and their corresponding bifurcation diagram are shown at right panels.}
\end{center}
\end{figure*}
\begin{figure}
\label{fig6}
\begin{center}
\includegraphics[height=280 pt]{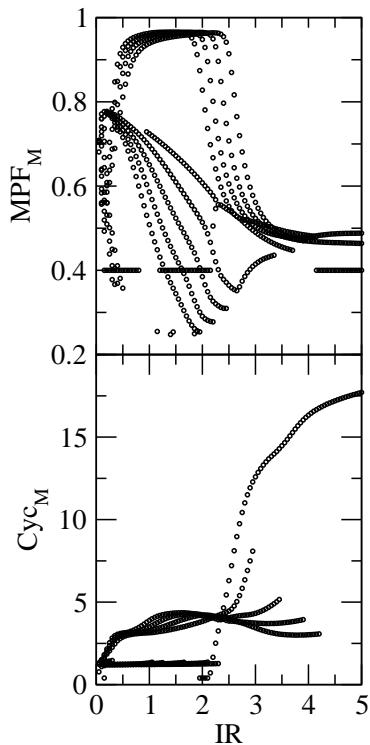}
\caption{Plot shows the impact of various IR dose on MPF maxima (at upper panel) as well as Cyclin maxima (at lower panel).}
\end{center}
\end{figure}

The perturbation induced by $p53$ through $IR$ to the cyclin via $p21$ clearly induces cyclin dynamics to various states shown by the bifurcation diagram (Fig. 4 right panels). We also notice that as one decreases or increases to cross over to sustain oscillation, the state just before it is chaotic regime. The emergence of oscillation death regime starts from $IR\rangle 3$ and then switch to sustain oscillation after sometime. This oscillation death regime corresponds to the collapse time due to strong sudden DNA damage. Once the DNA damage is recovered it comes back to sustain oscillation. If the $IR$ is very large then oscillation death regime is large enough that DNA damage can not be repaired back halting the cell cycle permanantly and goes to apoptosis.

\subsection{Dynamics of $MPF$ regulated by $p53$}

We present the temporal behaviour of $MPF$ regulated by $p53$ as a function of $IR$ (Fig. 5) which induce at different states in $MPF$ shown by bifurcation diagrams. The impact upon the $MPF$ due to $p53$ via $IR$ is not a direct phenomena but through $p21$ molecule in the network. Various studies reported that $p21$ directly interact with cyclin dependent kinases, which has very important role in the formation of maturation promoting factor (MPF). The interaction of $p21$ with $cdk$ leads to less availability of $cdk$ due to the formation of $MPF$. Moreover, various experimental results also reported that $p21$ directly interacts with $MPF$ \cite{bun,agu}. It is observed that a $IR=0$, the $MPF$ dynamics shows sustain oscillatory behaviour showing no impact of $p53$. Further, as $IR$ dose increases the oscillatory behaviour of $MPF$ is abruptly changed inducing different states of $MPF$ as we obtained in the case of cyclin. The increase in $IR$ induce different states oscillation death, period 1, 2, 3, chaotic and sustain oscillation regimes indicated by the bifurcation diagram for various $IR$ values. Moreover, as $IR$ increases the width of oscillation death \cite{bun} regime also increases and if $IR$ is not strong enough the DNA can able to repair back otherwise the system will go to apoptosis.

\subsection{Bifurcation in $MPF$ and Cyclin}

We study the regulation of cell cycle dynamics by $p53$ via $IR$. The maxima values of $MPF$ ($MPF_M$) and cyclin ($Cyc_M$) as a function of $IR$ are calculated for a range of time in the range $[0,50]$ hours (Fig. 6). It is observed that for low $IR$ dose, $MPF_M$ exhibits chaotic behaviour. Howeover, if $IR$ dose is comparatively high, $MPF_M$ becomes almost constant. If the value of $IR$ is moderate, period 1, 2, 3 etc are exhibited in the bifurcation diagram. This indicates that $MPF_M$ is $p53$ dependent via $IR$ and $p53$ controls the $MPF_M$ behaviour in the system. 

Similarly, one can also observe the $IR$ dependent maxima of cyclin $Cyc_M$ in the bifurcation diagram (Fig. 6 lower panel). The moderate values of $IR$ induce different periods in $Cyc_M$. Excess values of $IR$ shows different behaviour in $Cyc_M$. 

\section{Conclusion}
 
We study the way how $p53$, one of the largest hubs in cellular network, regulates and controls cell cycle dynamics. We studied the behaviour of different molecules which are highly involved in the checking of cell cycle at G2 phase driven by $p53$ via $IR$. The simulation results of the model provided us to understand the biological phenomenon and mechanism of cell cycle arrest due to DNA damage faced by the cell due to the irradiation. The results we got are closely in agreement with the previous experimental reports \cite{bun,bat}. Our study suggested that the temporal dynamics of molecular species involved in cell cycle, considered in the model, are controlled by $p53$. The role of $p21$ protein in the delay of G2 phase was considered as a cross-talk between $p53$ regulatory network and cell cycle. The sudden irradiation to the system with high dose induce collapse of the system due to DNA damage, leading to cell cycle arrest. The cell cycle is resumed again to normal situation by repairing back the DNA damage. Moreover, the time of recovery from cell cycle arrest and then resumption of oscillation depends on the amount of dose of $IR$ exposed to the system. 

During the process of regulation of cell cyle by $p53$ via $IR$ we observed the emergence of different periods (1, 2, 3 etc) in the bifurcation diagram of oscillatory dynamics of cell cycle variables ($MPF_M$ and $Cyc_M$) which may have various information of certain biological significance. Further, the dynamics of these variables switched to various states, namely, chaotic, oscillation death (stabilized state), bifurcating to various periods of oscillation and sustain oscillation states during the process of time evolution. These states could be the different phases of the variables to self-recover back to its normal condition from the sudden stress given to the system. However, how do these complicated states are used by the system dynamics when the system is perturbed need to be investigated further.

The study also demonstrates the mechanism of cell cycle arrest induced by perturbed $p53$ via $IR$ indicated by collapse of the oscillation (oscillation death) for certain interval of time ($\Delta t$). This collapse time is a function of strength of the perturbation imparted to the system. Our study shows that there is a minimum value of $IR=R_c$, below which the system comes back to its normal state, otherwise the system will go to apoptosis. Our findings will probably be useful for the further study on the impact of $p53$ on cell cycle checking at G2 phase and related dynamics.

\section*{Acknowledgments}
This work is financially supported by Department of Science and Technology (DST), New Delhi, India under sanction no. SB/S2/HEP-034/2012.

\begin{appendices}
\section{Tables of molecular species and biochemical reaction channels of the model}

\begin{table*}
\caption{Table 1: List of molecular species}
\begin{ruledtabular}
\begin{tabular}{|l|p{2.5cm}|p{5cm}|p{2cm}|}
\bf{ S.No.}    &   \bf{Species Name}    &    \bf{Description}           &  \bf{Notation}     \\ \hline
1.             &    $Cyclin$            & Unbounded Cyclin protein      &  $x_1$  \\ \hline
2.             &    $MPF$               & Maturation promotion factor   &  $x_2$  \\ \hline
3.             &    $Cyclin-Protease$   & Unbounded Cyclin Protease     &  $x_3$  \\ \hline
4.             &    $p53$               & Unbounded $p53$ protein       &  $x_4$ \\ \hline
5.             &    $Mdm2$              & Unbounded $Mdm2$ protein      &  $x_4$ \\ \hline
6.             &    $Mdm2\_p53$         & $Mdm2$ with $p53$ complex     &  $x_6$  \\ \hline
7.             &    $Mdm2\_mRNA$        & $Mdm2$ messenger $mRNA$       &  $x_7$  \\ \hline
8.             &    $ARF$               & Unbounded $ARF$ protein     	&  $x_8$  \\ \hline
9.             &    $ARF\_Mdm2$         & $ARF\_Mdm2$ complex           &  $x_9$  \\ \hline
10.            &    $IR$                & Irradiation		  	&  $x_{10}$  \\ \hline
11.            &    $DamDNA$            & Damaged DNA                   &  $x_{11}$  \\ \hline
12.            &    $p21$               & $p21$ protein                 &  $x_{12}$  \\ \hline
13.            &    $p21\_M$            & $p21$ and $M$  complex        &  $x_{13}$  \\ \hline
\end{tabular}
\end{ruledtabular}
\end{table*}

\begin{table*}
\begin{center}
{\bf Table 2 List of chemical reaction, rate constant and their values} 
\begin{tabular}{|l|p{2.5cm}|p{4.5cm}|p{3.8cm}|p{3cm}|p{1cm}|}
\hline \multicolumn{6}{}{}\\ \hline

\bf{ S.No.}    &   \bf{Biochemical reaction}    &    \bf{Description}           &  \bf{Rate Constant}   &  \bf{Values of Rate Constant}  &  \bf{Ref.}\\ \hline
1 & $\phi\stackrel{k_{1}}{\longrightarrow}x_1$ & Synthesis of Cyclin & $k_1$ & $0.000416667\times10^{-2}{sec}^{-1}$ &\cite{gold,nur,murr}\\ \hline
2 & $x_1\stackrel{k^*}{\longrightarrow}\phi$ & Decay of Cyclin & $k^* \langle x_1\rangle, where, k^*= \frac{k_2x_1x_3}{k_3+x_1}$ & $k_2=0.004166667{sec}^{-1}, k_3=0.02{sec}^{-1}$ & \cite{gold,stra}\\ \hline
3 & $x_1\stackrel{k_{4}}{\longrightarrow}\phi$ & Cyclin decay & $k_4 \langle x_1\rangle$ & $0.0000167{sec}^{-1}$ & \cite{gold}\\ \hline
4 & $\phi\stackrel{k^{**}}{\longrightarrow}x_2$ & Creation of MPF  & $k^{**},where, k^{**}=\frac{k_5(1-x_2)}{k_6+(1-x_2)},k_5=\frac{k_{14}x_1}{k_{13}+x_1} $ & $k_6=0.01,k_{13}=0.5,k_{14}=0.00{sec}^{-1}$ & \cite{bun,gold,rom} \\ \hline
5 & $x_2\stackrel{k^{***}}{\longrightarrow}\phi$ & Decay of MPF & $k^{***} \langle x_2\rangle, where, k^{***}=\frac{k_7x_2}{k_8x_2}$ & $k_7=0.0025{sec}^{-1}, k_8=0.01{sec}^{-1}$ & \cite{bun,murr2,rom} \\ \hline
6 & $x_2+x_{12}\stackrel{k_{31}}{\longrightarrow}x_{13}$ & Formation of $MPF\_p21$ complex & $k_{31}\langle x_2\rangle\langle x_{12}\rangle$  & $0.0001{mol}^{-1}{sec}^{-1}$ & \cite{bun,agu,har} \\ \hline
7 & $\phi\stackrel{k_{7}}{\longrightarrow}x_3$ & Activation of protease molecule & $k^{****}, where, k^{****}=\frac{k_9(1-x_3)}{k_{10}+(1-x_3)},k_9=x_2k_{15}$ & $k_{10}=0.01,k_{15}=0.001667$ & \cite{gold,mins1,bue} \\ \hline
8 & $x_3\stackrel{k^{*****}}{\longrightarrow}\phi$ & Inactivation of protease molecule & $k^{*****} \langle x_3\rangle, where, k^{*****}=\frac{k_{11}x_3}{k_{12}+x_3}$ & $k_{11}=0.0008333,k_{12}=0.01$ & \cite{gold,rom} \\ \hline
9 & $\phi\stackrel{k_{16}}{\longrightarrow}x_4$ & creation of p53 & $k_{16} $ & $0.078$ & \cite{fin,pro,jah} \\ \hline
10 & $x_4+x_5\stackrel{k_{17}}{\longrightarrow}x_6$ & synthesis of $p53\_MDM2$ complex & $k_{17} \langle x_4\rangle\langle x_5\rangle$ & $1.155\times 10^{-3}{mol}^{-1}{sec}^{-1}$ & \cite{pro,jah} \\ \hline
11 & $x_6\stackrel{k_{18}}{\longrightarrow}x_4+x_5$ & Dissociation of $p53\_MDM2$ complex & $k_{18} \langle x_6\rangle$ & $1.155\times 10^{-5}{sec}^{-1}$ & \cite{pro,moll,jah} \\ \hline
12 & $x_6\stackrel{k_{19}}{\longrightarrow}x_5$ & ubiquitination of p53 & $k_{19} \langle x_6\rangle$ & $8.25\times 10^{-4}{sec}^{-1}$ & \cite{fin,pro,jah}\\ \hline
13 & $x_4\stackrel{k_{20}}{\longrightarrow}x_4+x_7$ & creation of $MDM2\_mRNA$ & $k_{20} \langle x_4\rangle$ & $1.0\times 10^{-4}{sec}^{-1}$ & \cite{pro,lah,jah}\\ \hline
14 & $x_7\stackrel{k_{21}}{\longrightarrow}\phi$ & decay of $MDM2\_mRNA$ & $k_{21} \langle x_7\rangle$ & $1.0\times 10^{-4}{sec}^{-1}$ & \cite{pro,lah,jah}\\ \hline
15 & $x_7\stackrel{k_{22}}{\longrightarrow}x_5+x_7$ & synthesis of MDM2 & $k_{22} \langle x_7\rangle$ & $4.95\times 10^{-4}{sec}^{-1}$ & \cite{fin,pro,jah} \\ \hline
15 & $x_5\stackrel{k_{23}}{\longrightarrow}\phi$ & decay of MDM2 & $k_{23} \langle x_5\rangle$ & $4.33\times 10^{-4}{sec}^{-1}$ & \cite{fin,pro,jah} \\ \hline
16 & $x_{10}\stackrel{k_{24}}{\longrightarrow}x_{11}$ & creation of DNA damage & $k_{24} \langle x_{10}\rangle$ & $1.0{sec}^{-1}$ & \cite{vil,pro} \\ \hline
17 & $x_{11}\stackrel{k_{25}}{\longrightarrow}\phi$ & recovery of damaged DNA  & $k_{25}\langle x_{11}\rangle$ & $2.0\times 10^{-5}{sec}^{-1}$ & \cite{schu,pro} \\ \hline
18 & $x_{11}\stackrel{k_{26}}{\longrightarrow}x_8$ & Activation of ARF & $k_{26} \langle x_{11}\rangle$ & $3.3\times 10^{-5}{sec}^{-1}$ & \cite{pro}\\ \hline
19 & $x_5+x_8\stackrel{k_{27}}{\longrightarrow}x_9$ & synthesis of $MDM2\_ARF$ complex & $k_{27} \langle x_5\rangle\langle x_8\rangle$ & $0.01{mol}^{-1}{sec}^{-1}$ & \cite{pro,khan} \\ \hline
20 & $x_8\stackrel{k_{28}}{\longrightarrow}\phi$ & decay of ARF & $k_{28} \langle x_9\rangle\langle x_8\rangle$ & $0.001{sec}^{-1}$ & \cite{pro,kuo}\\ \hline
21 & $x_9\stackrel{k_{29}}{\longrightarrow}x_8$ & degradation of MDM2 & $k_{29}\langle x_9\rangle$ & $0.001{sec}^{-1}$ & \cite{pro,zhan} \\ \hline
22 & $x_4\stackrel{k_{30}}{\longrightarrow}x_4+x_{12}$ & synthesis of p21 & $k_{30} \langle x_4\rangle$ & $0.001{sec}^{-1}$ & \cite{dei,xio,deir} \\ \hline
23 & $x_2+x_{12}\stackrel{k_{30}}{\longrightarrow}x_{13}$ & synthesis of $p21\_MPF$ complex & $k_{31} \langle x_4\rangle$ & $0.0001{sec}^{-1}$ & \cite{dei,xio,deir} \\ \hline
24 & $x_{13}\stackrel{k_{32}}{\longrightarrow}x_{12}$ & dissociation of $p21\_MPF$ complex & $k_{32}\langle x_{13}\rangle$ & $0.002{sec}^{-1}$ & \cite{bun,agu,xio}\\ \hline
25 & $x_{12}\stackrel{k_{33}}{\longrightarrow}\phi$ & decay of p21 complex & $k_{33}\langle x_{12}\rangle$ & $0.005{sec}^{-1}$ & \cite{bun,xio}\\ \hline

\end{tabular}
\end{center}
\end{table*}

\newpage
\end{appendices}


\begin{thebibliography}{99}
\bibitem{lane} D. P. Lane, Nature {\bf 358}, 15 (1992).
\bibitem{men}S. M. Mendrysa, and M. E. Perry, Mol. Cell. Biol. {\bf 20}, 2023 (2000).
\bibitem{mic}D. Michael and M. Oren, Curr. Opin. Genet. Dev. {\bf 12}, 53 (2002).
\bibitem{bar}J. Bargonetti and J. J. Manfredi, Curr. Opin. Oncol. {\bf 14}, 86 (2002).
\bibitem{vog}B. Vogelstein, D. Lane, and A. J. Levine, Nature {\bf 408}, 307 (2000).
\bibitem{vou}K. H. Vousden, Cell {\bf 103}, 691 (2000).
\bibitem{mom}J. Momand, H. H. Wu, and G. Dasgupta, Gene {\bf 242}, 15 (2000).
\bibitem{mak}C. G. Maki, J. M. Huibregtse, and P. M. Howley, Cancer Res. {\bf 56}, 2649 (1996).
\bibitem{hon}R. Honda, H. Tanaka, and H. Yasuda, FEBS Lett. {\bf 420}, 25 (1997).
\bibitem{sch}M. Scheffner, J. M. Huibregtse, R. D. Vierstra, and P. M. Howley, Cell {\bf 75}, 495 (1993).
\bibitem{pom}J. Pomerantz, N. Schreiber-Agus, N. J. Liégeois, A. Silverman, L. Alland, L. Chin, J. Potes, K. Chen, I. Orlow, H. W. Lee, C. Cordon-Cardo, and R. A. DePinho, Cell {\bf 92}, 713 (1998).
\bibitem{hon1}R. Honda and H. Yasuda, EMBO J. {\bf 18}, 22, (1999).
\bibitem{mid}C. A. Midgley, J. M. Desterro, M. K. Saville, S. Howard, A. Sparks, R. T. Hay, and D. P. Lane, Oncogene {\bf 19}, 2312 (2000).
\bibitem{dei}W. S. el-Deiry, T. Tokino, V. E. Velculescu, D. B. Levy, R. Parsons, J. M. Trent, D. Lin, W. E. Mercer, K. W. Kinzler, and B. Vogelstein, Cell {\bf 75}, 817, (1993).
\bibitem{gal}A. L. Gariel, S. K. Radhakrishnan, Cancer Res. {\bf 65}, 3980 (2005).
\bibitem{bun}F. Bunz, A. Dutriaux, C. Lengauer, T. Waldman, S. Zhou, J. P. Brown, J. M. Sedivy, K. W. Kinzler, and B. Vogelstein, Science 
{\bf 282}, 1497, 1998.
\bibitem{bat}S. Bates, K. M. Ryan, A. C. Phillips, and K. H. Vousden, Oncogene {\bf 17}, 1691 (1998).
\bibitem{agu}B. D. Aguda, Proc. Natl. Acad. Sci. USA, {\bf 96}, 11352 (1999).
\bibitem{har}J. W. Harper, G. R. Adami, N. Wei, K. Keyomarsi, and S. J. Elledge, Cell {\bf 75}, 805 (1993).
\bibitem{apre}O. Aprelikova, Y. Xiong, and E. T. Liu, J. Biol. Chem. {\bf 270}, 18195 (1995).
\bibitem{fun}J. O. Funk, S. Waga, J. B. Harry, E. Espling, B. Stillman, and D. A. Galloway, Genes Dev. {\bf 11}, 2090 (1997).
\bibitem{sor}G. Soria, J. Speroni, O. L. Podhajcer, C. Prives, and V. Gottifredi, J. Cell Sci. {\bf 121}, 3271 (2008). 
\bibitem{res}D. Resnitzky and S. I. Reed, Mol. Cell. Biol. {\bf 15}, 3463 (1995).
\bibitem{hin}P. W. Hinds and R. A. Weinberg, Curr. Opin. Genet. Dev. {\bf 4}, 135 (1994).
\bibitem{lam}E. W. Lam and N. B. La Thangue, Curr. Opin. Cell Biol. {\bf 6}, 859 (1994).
\bibitem{pro}C. J. Proctor and D. A. Gray, BMC Syst. Biol.{\bf 2}, 75 (2008).
\bibitem{gold}A. Goldbeter, Proc. Natl. Acad. Sci. U. S. A. {\bf 88}, 9107 (1991).
\bibitem{moll}U. M. Moll and O. Petrenko, Mol. Cancer Res. {\bf 1}, 1001 (2003).
\bibitem{jah} M.J. Alam, G.R. Devi, Ravins, R. Ishrat, S.M. Agarwal, and R.K.B. Singh, Mol. BioSyst. {\bf 9}, 508 (2013).
\bibitem{fin}C. A. Finlay, Mol. Cell. Biol. {\bf 13}, 301 (1993).
\bibitem{lah}G. Lahav, N. Rosenfeld, A. Sigal, N. Geva-Zatorsky, A. J. Levine, M. B. Elowitz, and U. Alon, Nat. Genet. {\bf 36}, 147 {2004}.
\bibitem{vil}M. M. Vilenchik and A. G. Knudson, Proc. Natl. Acad. Sci. U. S. A. {\bf 100}, 12871 {2003}.
\bibitem{schu}L. B. Schultz, N. H. Chehab, A. Malikzay, and T. D. Halazonetis, J. Cell Biol. {\bf 151}, 1381 (2000).
\bibitem{khan}S. Khan, C. Guevara, G. Fujii, and D. Parry, Oncogene {\bf 23}, 6040 (2004).
\bibitem{kuo}M.L. Kuo, W. den Besten, D. Bertwistle, M. F. Roussel, and C. J. Sherr, Genes Dev. {\bf 18}, 1862 (2004).
\bibitem{zhan}Y. Zhang, Y. Xiong, and W. G. Yarbrough, Cell {\bf 92}, 725 (1998).
\bibitem{xio}Y. Xiong, G. J. Hannon, H. Zhang, D. Casso, R. Kobayashi, and D. Beach, Nature {\bf 366}, 701 (1993).
\bibitem{deir}W. S. el-Deiry, J. W. Harper, P. M. O’Connor, V. E. Velculescu, C. E. Canman, J. Jackman, J. A. Pietenpol, M. Burrell, D. E. Hill, and Y. Wang, Cancer Res. {\bf 54}, 1169 (1994).
\bibitem{wag}S. Waga, G. J. Hannon, D. Beach, and B. Stillman, Nature {\bf 369}, 574 (1994).
\bibitem{morl}A. O. Morla, G. Draetta, D. Beach, and J. Y. Wang, Cell {\bf 58}, 193 (1989).
\bibitem{stra}U. Strausfeld, J. C. Labbé, D. Fesquet, J. C. Cavadore, A. Picard, K. Sadhu, P. Russell, and M. Dorée, Nature, {\bf 351}, 242 (1991).
\bibitem{murr2}A. W. Murray, M. J. Solomon, and M. W. Kirschner, Nature {\bf 339}, 280 (1989).
\bibitem{mins}J. Minshull, J. Pines, R. Golsteyn, N. Standart, S. Mackie, A. Colman, J. Blow, J. V. Ruderman, M. Wu, and T. Hunt, J. Cell Sci. Suppl. {\bf 12}, 77 (1989).
\bibitem{rom}P. C. Romond, M. Rustici, D. Gonze, and A. Goldbeter, Ann. N. Y. Acad. Sci. {\bf 879}, 180 (1999).
\bibitem{mins1}J. Minshull, R. Golsteyn, C. S. Hill, and T. Hunt, EMBO J., {\bf 9}, 2865 (1990).
\bibitem{bue}B. Buendia, P. R. Clarke, M. A. Félix, E. Karsenti, D. Leiss, and F. Verde, Cold Spring Harb. Symp. Quant. Biol. {\bf 56}, 523 (1991).
\bibitem{pre}W. H. Press, S. A. Teukolsky, W. T. Vetterling, and B. P. Flannery, Numerical Recipe in Fortran. Cambridge University Press (1992).
\bibitem{nur}P. Nurse, Nature {\bf 344}, 503 (1990).
\bibitem{murr}A. W. Murray and M. W. Kirschner, Science, {\bf 246}, 614 (1989).
\end{thebibliography}
\end{document}